\documentstyle[aps]{revtex}

\newcommand{\beq}{\begin{equation}}
\newcommand{\eeq}[1]{\label{#1}\end{equation}}
\newcommand{\beqn}{\begin{eqnarray}}
\newcommand{\eeqn}[1]{\label{#1}\end{eqnarray}}
\newcommand{\ba}{\begin{array}}
\newcommand{\ea}{\end{array}}

\newcommand{\As}{\not\!\! A}
\newcommand{\Bs}{\not\!\! B}
\newcommand{\ds}{\!\!\not\!\partial}
\newcommand{\as}{\not\!\! a}
\newcommand{\bs}{\not\!\! b}

\newcommand{\D}{{\cal{D}}}

\newcommand{\Z}{{\cal Z}}

\newcommand{\bp}{\psi^{\dagger}}

\begin{document}

\begin{center}
\hfill IC/97/27 \\
\hfill La Plata-Th 97/04 \\

\vspace{0.3in}
{\Large\bf Spinons and parafermions in fermion cosets}
\\[.3in]
{\large Daniel C. Cabra\footnote{Investigador CONICET, Argentina.\\
On leave from Universidad Nacional de La Plata,
Argentina.\\
{\it Talk given at the International
Seminar ``Supersymmetry and Quantum Field Theory" dedicated to the memory
of D.V.Volkov, Kharkov, Ukraine, 5-7 jan. 1997}}}\\
\bigskip {\it International Centre for Theoretical Physics, Strada Costiera
11, 34100 Trieste, Italy}\\
\end{center}

\vspace{.3cm}

\begin{quotation}
We introduce a set of gauge invariant fermion fields in fermionic coset models 
and show that they play a very central role in the description of several 
Conformal Field 
Theories (CFT's). In particular we discuss the explicit 
realization of primaries and their OPE in unitary minimal models,
parafermion fields in $Z_k$ CFT's and that 
of spinon fields in $SU(N)_k,~k=1$ Wess-Zumino-Witten models (WZW) theories. 
The higher level case 
($k>1$) will be briefly discussed. Possible applications to QHE systems and 
spin-ladder systems are addressed.
\end{quotation}

\vspace{1cm}

\noindent {\large \bf Introduction}

\vspace{.5cm}

\setcounter{footnote}{0}

The purpose of this talk is to briefly review the fermionic coset
description of some particular CFT's, whose relevance has shown up in
particular in the construction of the primary operators in the minimal
unitary models and the $Z_k$ parafermion models, as well as in the
identification of quasiparticle operators in $SU(N)_k$ WZW models
\cite{Wi,KZ}.  We will particularly emphasize the realization of
order-disorder (OD) algebras as well as the realization of quasi-particle
operators with generalized statistics, where the gauge invariant fermion
fields play a central role.

We want to point out that we will only briefly comment on some of the 
basics of CFT which are relevant for 
the discussion we want to pursue. For more details 
see for example ref.\cite{Ginsparg}.

The so called coset construction was introduced by Goddard, Kent and Olive 
\cite{GKO}
as a way to obtain CFT's with Virasoro central extension less than unity, and 
can be 
used to build up many interesting conformal field theories \cite{Ginsparg}.
Lagrangian realizations of these coset theories have been presented, 
both in the bosonic
formulation \cite{GK}, in terms of gauged WZW models 
and in the fermionic versions \cite{BRS}, in terms of constrained (gauged) 
fermions.

Here we will use the fermionic description of coset models, and will 
discuss essentially three cases: the models in the so called 
minimal unitary or FQS series \cite{FQS}, the $Z_k$ parafermion models 
\cite{FK} and 
the $SU(N)_k$ WZW models \cite{KZ,NS} (from the point of view of fermion 
cosets). However, the present approach can be applied to the 
study of arbitrary coset models.
(Details on the computations that lead to the results reviewed here are 
contained in refs. \cite{CR1,CM,CR,CMR}).

In the case of minimal models we have shown that all the primaries 
can be constructed as certain fermion bilinears, and in the case of the Ising 
model, (the first model in the minimal unitary series), we have shown  that the
role of the gauge invariant fermions is crucial in the identification of the 
OD fields. This is the first case in which the relevance of 
gauge invariant fermions has shown up \cite{CR1}.

In ref.\cite{CM} we have shown how the construction of the OD operators
in \cite{CR1}
can be generalized to construct all the primaries in $Z_k$ parafermion models,
and we used this construction to study the ``thermal" perturbation of the 
system.

More recently \cite{CR}, we have shown that the gauge invariant fermions
realize in a natural way the so called ``spinon" fields, which were shown to 
play 
a crucial role in the quasiparticle description of the Hilbert space of the 
$SU(N)_1$ WZW theories, as motivated by the underlying Yangian 
symmetry \cite{BLS}.
(The extension to higher level cases is in progress \cite{CMR}).
As a byproduct, we have also shown how to factorize the WZW primaries in 
holomorphic and anti-holomorphic parts, a problem that has received some 
attention recently \cite{BCR}.

All these results clearly suggest that the gauge invariant fermions (gif's) 
play an
important role in the description of CFT's: in the case of $SU(N)_k$ WZW 
models, these gif's are precisely the quasiparticle operators, 
and both in the minimal unitary models and in the $Z_k$ parafermion models, 
all the primaries are composites (bound states) of these gif's.
Hence, there is a good chance that the construction of quasi-particle 
operators in terms of gif's
pursued in refs.~\cite{CR,CMR} could be extended to other CFT's. This 
construction
could be of some interest in connection with the so called quasiparticle (or
fermionic) representation of the characters for CFT's \cite{MC}.
 
Another interesting point is that the quasiparticle operators constructed in 
ref.\cite{CMR}
for the higher level cases satisfy non-abelian braiding relations, and hence 
could
play a role in the description of some QHE systems, such as the so called 
Haldane-Rezayi
and Pfaffian states  \cite{MR}, where the elementary quasiparticles have 
non-abelian statistics.

Finally we want to mention the so called spin-ladder systems 
(see ref.\cite{DR}, 
and references 
therein), which in the low energy regime can be studied as certain 
WZW models with interactions. Our approach could be also useful in the 
study of 
these systems,
where many interesting effects have a non-perturbative origin.

\vspace{.5cm}

\noindent {\large \bf An example: $SU(N)_1$ WZW theory as a fermion coset}

\vspace{.5cm}

Let us first set up our conventions and describe the approach in a simple 
example,
such as the fermionic coset representation of the $SU(N)_1$ WZW 
theory \cite{NS}.

The idea is to describe this model as a fermionic coset $U(N)/U(1)$, which is 
constructed starting with $N$ free massless Dirac fermions and freezing the 
$U(1)$ charge by imposing

\beq
j_{\mu}|phys>=0,
\eeq{1}
where $j_{\mu}$ is the $U(1)$ fermionic current.

This is achieved in the path-integral by introducing a $\delta$-functional
as

\beq
\delta[j_{\mu}]=\int Da_{\mu} exp\left(-\int d^2 x \psi^{\dagger} 
\gamma_{\mu} \psi
a_{\mu}\right).
\eeq{2}

The partition function of the $U(N)/U(1)$ model is then given by

\beq 
{\cal Z}_{U(N)/U(1)}=\int D \psi^{\dagger} D\psi Da_{\mu}
exp\left(-\int d^2 x  \psi^{\dagger} (i\ds +i\as) \psi \right),
\eeq{3}
and is equivalent to the partition function of the $SU(N)_1$ WZW 
model in the sense that correlators of corresponding fields in the two 
theories coincide \cite{NS}.

In order to write the partition function in a more manageable form,
we perform the following change of variables \cite{FiPol}

\beq
\begin{array}{ll}
a=i(\bar{\partial}u)u^{-1} &
\bar{a}= i(\partial\bar{u})\bar{u}^{-1},  \\
\psi_1=u\chi_1 & \bp_2=\chi_2^{\dagger} u^{-1},\\
\psi_2=\bar{u}\chi_2 & \bp_1=\chi_1^{\dagger} \bar{u}^{-1}.
\end{array}
\eeq{4}
where
$\partial \equiv \frac{\partial}{\partial z}$,
$\bar{\partial} \equiv \frac{\partial}{\partial \bar z}$
and $\psi=\left(
\begin{array}{c}
\psi_1 \\ \psi_2
\end{array}
\right)$.
The fields $u$ and $\bar u$ are parametrized in terms of real scalar
fields as
$h=\exp(-\phi -i\eta)$ and
$\bar h=\exp(\phi -i\eta)$.

Taking into account the gauge fixing procedure and the Jacobians
associated to (\ref{4}) \cite{FiPol} one arrives at
the desired decoupled form for the partition function:
\beq
\Z=\Z_{ff}\Z_{fb}\Z_{gh},
\eeq{5}
where
\beqn
\Z_{ff} &=& \int \D\chi^{\dagger} \D\chi \exp( -\frac{1}{\pi}
\int(\chi^{\dagger}_2 \bar{\partial} \chi_1 +
\chi^{\dagger}_1 \partial \chi_2)d^2x), \nonumber \\
\Z_{fb} &=& \int \D\phi \exp(\frac{N}{2\pi}\int \phi \Delta \phi d^2x).
\eeqn{6}
(The explicit form of the ghost partition function is inessential in what 
follows).

Notice that, although the partition function of the theory is completely
decoupled, BRST quantization conditions connect the different sectors in 
order to ensure unitarity \cite{GK}.

The central charge is now easily evaluated as the sum of three
independent
contributions coming from the different sectors, $c_{ff}=N$,
$c_{fb}=1$ and $c_{gh}=-2$, thus giving $c=N-1$
which coincides with the central charge of the $SU(N)_1$ WZW model.
Similarly, conformal dimensions of primaries can be evaluated using this 
decoupled picture.

\vspace{.5cm}

More general models, as for example the $SU(2)_k$ WZW models,
can be also represented as fermionic cosets by making use of 
the general equivalence \cite{NS}
\beq
U(2k)/\left(U(1)\times SU(k)_2\right).
\eeq{7}
In this case, in addition to the constraint implemented by the abelian gauge 
field 
$a_{\mu}$, we have to introduce another constraint associated with
the $SU(k)$ currents. This constraint will be implemented by a
non-abelian gauge field in the Lie algebra of $SU(k)$.
The $SU(k)$ gauge field is traded for a $SU(k)$ WZW field through a
decoupling transformation similar to that of eq.(\ref{4}).

Using the approach described above we have studied the following cases:

\vspace{.5cm}

\noindent {\large \bf i) Minimal unitary models}

\vspace{.5cm}

These models can be represented as the cosets \cite{GKO}

\beq
\left(SU(2)_k\times SU(2)_1\right)/SU(2)_{k+1},
\eeq{8}
and the Virasoro central charges lie in the FQS series \cite{FQS}
\beq
c=1-6/\left((k+2)(k+3)\right).
\eeq{9}
By making use of eq.(\ref{7}), one is led to 
make the identification of the coset (\ref{8}) with the fermion coset
\beq
\left[\left(\frac{U(2k)}{SU(k)_2\times U(1)}\right)\times 
\left(\frac{U(2)}{U(1)}\right)\right]/SU(2)_{k+1},
\eeq{10}
whose Lagrangian is given by

\beq
{\cal L}=\psi^{\dagger} \left((\ds +\as +\Bs +\As \right) \psi+
\bar \chi^{\dagger} \left(\ds +\bs +\As \right) \chi,
\eeq{11}
where all the internal indices are supressed. (The fermions $\psi$ ($\chi$)
transform in the fundamental representation of $U(2k)$ ($U(2)$)).
The gauge fields $a_{\mu}$, $b_{\mu}$,$B_{\mu}$,$A_{\mu}$, in the Lagrangian 
(\ref{11}),
implement respectively the $U(1)$, $U(1)$, $SU(k)_2$ and $SU(2)_{k+1}$. 
(The subindices refer to the central charge of the constrained affine 
currents).

Within this approach one can show that the central charge of these fermionic
coset models is given by eq.(\ref{9}) \cite{BRS}, all the primaries can be
constructed in terms of fermion bilinears \cite{CR1}, and the correlators 
can be evaluated in terms of known results on WZW theories \cite{FZ2}.

The point we want to make here is that the fermionic description is more 
suitable
than the bosonic one 
regarding the construction of primary fields and their operator product 
algebra.
In particular we can identify in a natural way the OD fields as we will 
show in the 
examples bellow.
There is one interesting example that has been completely described
within this scheme, which corresponds to the Ising model ($c=1/2$, $k=1$ in 
eq.(\ref{9})).
In this case, not only the primary operators, (i.e. the fields with
their dimensions in the Kac table), but also the OD
algebra has been realized using the gauge invariant fermions \cite{CR1}.

\vspace{.2cm}

{\bf Gauge invariant fermions}

\vspace{.2cm}

Let us introduce the gauge invariant fermion fields 
of the theory defined by eq.(\ref{11}), for $k=1$,
\beqn
\hat\psi^i(x)&=&e^{-i\int^\infty_xdz_\mu a_\mu}
\left(Pe^{-i\int^\infty_xdz^\mu A_\mu}\right)_{ij}\psi^j(x) \label{ca}\\
\hat\chi^i(x)&=&e^{-i\int^\infty_xdz_\mu b_\mu}
\left(Pe^{-i\int^\infty_xdz^\mu A_\mu}\right)_{ij}\chi^j(x).
\eeqn{12}
(Note that for
$k=1$, the $B$ field is not present)

In terms of these fields, the spin ($\sigma$) and disorder ($\mu$) fields read
\beq
\sigma(x)=\hat\psi^\dagger\hat\psi+\hat\chi^\dagger
\hat\chi, \ \ \ \ and \ \ \ \mu(x)=\hat\psi^\dagger\hat\chi+\hat\chi^\dagger
\hat\psi.
\eeq{14}

Using the decoupling formulation one can show that they have the correct 
dimensions (i.e. $h=\bar h= 1/16$), and satisfy the OD algebra
\cite{ST}
\beq
\sigma(z_1)\mu(z_2)=e^{i\pi \Theta(z_1-z_2)}\mu(z_2)\sigma(z_1),
\eeq{15}
where $\Theta(z)$ stands for the Heaviside function.

We will explain how the OD algebra arises in a simpler example later on, 
but let
us point out that while the Schwinger line integrals associated with the 
non-abelian gauge field $A$ cancel out in the bilinears (\ref{14}),
those associated with the abelian gauge fields $a$ and $b$ will not, 
and this fact will give rise to the OD algebra.

\vspace{.5cm}

\noindent {\large \bf ii) $Z_k$ parafermion models}

\vspace{.5cm}

These models are realized as the cosets
\beq 
SU(2)_k/U(1),
\eeq{16}
and the Virasoro central charge is given  by
$c=2(k-1)/(k+2)$.

In terms of fermionic cosets they correspond to

\beq
\left(\frac{U(2k)}{U(1)\times SU(k)_2}\right)/U(1),
\eeq{18}

In order to construct the primary fields we have proceeded
in two steps: first we have used the identification of the primary fields
of the $SU(2)_k$ WZW in terms of fermion bilinears made in ref.\cite{NS},
and then, rewritting these bilinears in terms of gif's
we have shown that they correspond to OD fields.
Other primaries of the $Z_k$ models, such as parafermion currents, 
were also constructed as suitable composites of gauge invariant fermions.

\vspace{.2cm}

\noindent {\bf $SU(2)_k$ primaries}

\vspace{.2cm}

The Lagrangian of the fermionic description of the $SU(2)_k$ WZW  model is 
given by
\beq
{\cal L}=\psi^{\dagger} \left((\ds +\as +\Bs \right) \psi,
\eeq{19}
where $a_{\mu}$ and $B_{\mu}$ are $U(1)$ and $SU(k)$ gauge fields, 
implementing the respective constraints .

The fundamental field $g$ and its adjoint $g^{\dagger}$ of the
bosonic $SU(2)_k$
WZW theory are represented in terms of fermions by the bosonization
formulae \cite{NS}
\beq
g^{ij}=\psi_2^i\psi_2^{j\dagger}, \ \ \ \ \ \ 
g^{ij\dagger}=\psi_1^i\psi_1^{j\dagger} .
\eeq{20}
All other integrable representations can be constructed as
symmetrized normal ordered products of these fundamental fields.
Note that these fields are invariant under gauge transformations, both in 
$U(1)$ 
and $SU(k)$, as they should be in order to correspond to physical operators. 

The theory which we are interested in has one more $U(1)$ constraint
(see eq.(\ref{18})), which we implement by adding to the Lagrangian (\ref{19})
the term
\beq
\Delta {\it L}=\psi^{\dagger} \bs^3 t^3\psi,
\eeq{21}
where $t^3$ is an $SU(2)$ generator.
This term (after functional integration over $b_{\mu}^3$) 
implements the additional $U(1)$ constraint as follows from  
eqs.(\ref{1}),(\ref{2}).
It is easy to show, using the approach described above, 
that the central charge of the resulting model is the correct one.

The fields in (\ref{20}) vary under gauge transformations 
associated with the
new gauge field, $b_{\mu}^3$, introduced in eq.(\ref{21}).  In order to
ensure invariance also under these transformations, we will define
the gauge invariant fermions as

\beq
\hat{\psi}=e^{-i\int_x^{\infty} dz^{\mu}b_{\mu}^3}t^3\psi .
\eeq{25}

In \cite{CM} we have shown that all the $Z_k$ primaries 
can be built up from these fields. Here we will only discuss an example 
in some detail, in order to show how OD algebras are realized in 
terms of gifs.

Using (\ref{25}) we can construct the gauge invariant version of the
$g$-field and its adjoint in eq.(\ref{20}) \cite{CR1}
\beq
{\hat g}_{ij}={\hat{\psi}}_2^i{\hat{\psi}}_2^{j\dagger},
\ \ \ \ \ \ \ \ \
{\hat g}_{ij}^{\dagger}={\hat{\psi}}_1^i{\hat{\psi}}_1^{j\dagger}.
\eeq{26}
Let us consider the composites
\beq
\sigma_1\equiv  {\hat g }_{1,1},
\ \ \ \ \ \ \ \ \ 
\mu_1\equiv  {\hat g }^{\dagger}_{2,1} .
\eeq{27}
In the decoupled picture these fields can be rewritten as
\beqn
\sigma_1&\equiv & {\hat g}_{1,1}=
:\chi_2^{1\alpha} {\tilde U}^{-1\ \alpha\beta}
\chi_2^{\dagger 1\beta}: :e^{2\phi_a}::e^{2\phi_b}:
,
\nonumber \\
\mu_1&\equiv & {\hat g}^{\dagger}_{2,1}=
:\chi_1^{2\alpha} {\tilde U}^{\alpha\beta}
\chi_1^{\dagger 1\beta}:
:e^{-2\phi_a}::e^{\varphi_b-{\bar \varphi}_b}:.
\eeqn{27'}
Here $\tilde U$ is the $SU(k)$ WZW field that parametrizes the field $B_{\mu}$,
and $\phi_a$, $\phi_b$ are the $U(1)$ boson fields that parametrize $a_{\mu}$
and $b_{\mu}$ respectively, and $\varphi_b$ and $\bar{\varphi}_b$ 
are the chiral (holomorphic and anti-holomorphic) components of the free
boson $\phi_b$. 

The dimensions of these fields are easily evaluated in the decoupled picture 
and are given by
\beq
h=\bar{h}=\frac{k-1}{2k(k+2)}
\eeq{28}
and it can be shown that they satisfy the OD algebra \cite{CR1}
\beq
\sigma_1(x_1) \mu_1(x_2) =e^{\frac{i2\pi}{k}\Theta(x_1-x_2)}
\mu_1(x_2)\sigma_1(x_1).
\eeq{29}
Eqs. (\ref{28}) and (\ref{29}) lead one to identify the fields $\sigma$
and $\mu$ defined in (\ref{27}) with the order and disorder operators in
the $Z_k$ parafermion theory.

Let us stress that the OD algebra has its origin in the particular way in 
which
the holomorphic components of the free boson $\phi_b$ are combined
in eq.(\ref{27'}). This, in turn, is a consequence of the use of the gauge
invariant fermions (\ref{25}). Indeed, one can easily
check, using the cannonical commutation rules for $\phi_b$ that
\beq
:e^{2\phi_b(x_1)}::e^{(-\varphi_b+{\bar{\varphi}}_b)(x_2)}:
=e^{\frac{i2\pi}{k}\Theta(x_1-x_2)}
:e^{(-\varphi_b+{\bar{\varphi}}_b)(x_2)}::e^{2\phi_b(x_1)}:   ,
\eeq{27''}
being the other factors commuting.

This construction can be generalized to all other $Z_k$ primaries and hence,
having identified all the primaries one can pursue the study of 
perturbations in the Lagrangian approach.
We have studied the ``thermal" perturbation of $Z_k$ models within the
present approach and reduced the problem to the study of interacting WZW 
fields.
In this scheme, two fixed points of the perturbed system were identified 
\cite{CM}.

\vspace{.5cm}

\noindent {\large \bf iii) $SU(N)_k$ WZW models: gif's, spinons and 
holomorphic factorization}

\vspace{.5cm}

All we have done so far is to study bilinears of the gauge invariant fermions.
In this last section we will study the properties of the gauge invariant 
fermions 
themselves and show their relevance in the so called ``spinon" description
of WZW models \cite{BLS}, as described in ref.\cite{CR,CMR}.

Let us then define the gauge invariant fermion fields \cite{CR} corresponding
to the $SU(N)_k$ theory similarly as in (\ref{ca}), (with $A_{\mu}$ 
replaced by $B_{\mu}$),
that by construction will create the physical excitations, and let us study 
their properties.
For simplicity we will discuss the level one case, although 
similar considerations apply to the higher level case \cite{CMR}.

In the decoupled picture, the gifs are given by

\beq
\begin{array}{ll}
\hat{\psi_1}^i(z)
=e^{-\varphi(z)}\chi_1^i(z)
& \hat{\psi_1}^{i \dagger}(\bar z) =
\chi_1^{i \dagger}(\bar z) e^{-\bar{\varphi}(\bar z)}
\\
\hat{\psi}_2^i(\bar z) =
e^{\bar{\varphi}(\bar z) }\chi_2^i(\bar z)
& \hat{\psi_2}^{i \dagger}(z)
=\chi_2^{i \dagger}(z) e^{\varphi(z)}
\end{array}
\eeq{31}
where $\varphi (z)$ and $\bar{\varphi}$ 
are the chiral (holomorphic and anti-holomorphic) components of the free
boson $\phi$. 
This fact together with the equation of motion of the free fermions
$\chi$ ensures
that $\hat{\psi}^i_1$ and $\hat{\psi}^{i\dagger}_2$ ($\hat{\psi}^i_2$
and
$\hat{\psi}^{i\dagger}_1$)
are holomorphic (anti-holomorphic).
The conformal dimensions are given by $((N-1)/2N,0)$ and $(0,(N-1)/2N)$
respectively, thus suggesting that they correspond to the ``halves" of the WZW 
primary $g(z,\bar z)$.

In fact, it can be shown by the following sequence of identities
\beq
g^{ij}(z, \bar z)=
\psi_2^i\psi_2^{j\dagger}=
{\hat{\psi}}_2^i(\bar z){\hat{\psi}}_2^{j\dagger}(z),
\eeq{32}
that the gifs are the holomorphic factors of the WZW primary.

The second important property satisfied by these operators is their OPE 
which allows us to identify them as the ``spinon" operators of ref.\cite{BLS}:

\beq
\hat{\psi_1}^i(z) \hat{\psi_1}^j(w)=
\frac{1}{(z-w)^{1/N}}{\cal A}(:\chi_1^i(w)\chi_1^j(w):)
:\exp 2\varphi(w):+\dots  = \frac{1}{(z-w)^{1/N}} \Phi_2^{ij}(w) + \dots,
\eeq{33}
where ${\cal A}$ stands for antisymmetrization and $\Phi_2^{ij}$ is the WZW 
primary with dimension $(N-2)/N$.

As explained in the third reference of
\cite{BLS} the branch cut in the OPE singularity implies that
the excitations created by the spinon fields satisfy generalized statistics
with ``statistical angle" $\theta =\pi /N$.
The chiral Fock space of the $SU(N)_1$ CFT can be constructed in terms
of the modes of these fields and this space can be classified into multiplets
corresponding to the irreducible representations of the Yangian algebra 
$Y(sl_N)$ \cite{BLS}.

In the case of higher levels, the same construction can be done, and in this
case,
the excitations created by the gauge invariant fermions satisfy non-abelian
braiding statistics. The holomorphic factorization of the WZW fields in terms
of gifs can also be proved. The results of this investigation will appear 
elsewhere \cite{CMR}.

\vspace{.2cm}

{\it  Acknowledgements}: I would like to thank to my collaborators
E. Moreno, G. Rossini and K.D.Rothe with whom the work presented
in this talk has been developed. This work was supported in part by CONICET 
and Fundaci\'on Antorchas, Argentina.

\end{document}